\PassOptionsToPackage{dvipsnames,svgnames,x11names}{xcolor}
\documentclass[article,copyright,creativecommons]{eptcs}
\providecommand{\event}{Non-Classical Logics. Theory and Applications 2026} 

\usepackage{iftex}

\ifpdf
  \usepackage{underscore}
  \usepackage[T1]{fontenc}
\else
  \usepackage{breakurl}
\fi

\usepackage[dvipsnames,svgnames,x11names]{xcolor}
\newcommand{\todo}[2][]{ }
\usepackage{amsthm}
\usepackage[normalem]{ulem} 

\usepackage{framed}

\usepackage{amssymb,mathtools}
\usepackage{enumitem}
\usepackage{tikz}

\usepackage{tabularx,array}
\usepackage{comment}
\newcolumntype{L}{>{\raggedright\arraybackslash}X}

\theoremstyle{plain}
\newtheorem{theorem}{Theorem}[section]   
\newtheorem{lemma}[theorem]{Lemma}       

\newtheorem{proposition}[theorem]{Proposition}

\theoremstyle{definition}
\newtheorem{definition}[theorem]{Definition}

\theoremstyle{remark}
\newtheorem{remark}[theorem]{Remark}

\newcommand{\et}{\mathbin{\wedge}}

\allowdisplaybreaks

\newcommand{\CL}{\mathsf{CL}}

\newcommand{\CLoN}{\mathsf{CLoN}}
\newcommand{\CLuN}{\mathsf{CLuN}}
\newcommand{\CLaN}{\mathsf{CLaN}}
\newcommand{\CLauN}{\mathsf{CLauN}}
\newcommand{\CLoaN}{\mathsf{CLoaN}}
\newcommand{\CLouN}{\mathsf{CLouN}}
\newcommand{\uneg}{\mathord{\sim_u}}
\newcommand{\aneg}{\mathord{\sim_a}}
\newcommand{\oneg}{\mathord{\sim_o}}
\newcommand{\cneg}{\mathord{\neg_c}}

\newcommand{\Forms}{\mathsf{Form}}
\newcommand{\Var}{\mathsf{Var}}
\newcommand{\defeq}{\mathrel{:=}}

\newcommand{\Val}{\mathsf{Val}}

\newcommand{\Goal}[1]{}

\hypersetup{
  colorlinks=true,
  urlcolor=RoyalBlue,
  citecolor=RoyalBlue,
  linkcolor=RoyalBlue,
}

\begin{document}

\title{Paraclassical Logics with Two Negations}

\author{Pawel Pawlowski
	\institute{Centre for Logic and Philosophy of Science\\
		Ghent University, Ghent, Belgium}
	\and
	Pietro Vigiani
	\institute{Institute of Computer Science\\\smallskip
		Czech Academy of Sciences, Prague, Czech Republic\\
		Centre for Logic and Philosophy of Science\\
		Ghent University, Ghent, Belgium}
}
\def\titlerunning{Paraclassical Logics with Two Negations}
\def\authorrunning{P. Pawlowski \& P. Vigiani}

\maketitle

\begin{abstract}
	We study three paraclassical propositional logics with two
	negations, \(\CLoaN\), \(\CLouN\), and \(\CLauN\), obtained by pairwise combining the one-negation systems \(\CLuN\),
	\(\CLaN\), and \(\CLoN\).  $\CLuN$'s negation \(\uneg\) is glut-tolerant and
	paraconsistent, $\CLaN$'s negation \(\aneg\) is gap-tolerant and paracomplete, and $\CLoN$'s negation $\oneg$ is both paraconsistent and paracomplete. We
	provide sound and complete two-valued non-deterministic semantics for
	each of the two-negations systems. We then focus on \(\CLauN\), the system containing the
	primitive dual pair \((\aneg,\uneg)\). In this setting, we show deductive relationships between different negations expressible in $\CLauN$; we show two diferent ways to recapture classical logic by means of translations, locally and globally; and we show a duality results for the multiple conclusion consequence relation of the logic.
\end{abstract}
\section{Introduction}

Negation is one of the most philosophically interesting connective, and its understanding is a point of divergence of many non-classical logics. For the purpose of our paper, two natural weakenings of the properties of classical negation are especially interesting. One option is to give up \emph{ex contradictione quodlibet}, so that a formula and its negation no
longer jointly entail everything. The resulting negation is then paraconsistent, which on the semantics side usually entails that such a logic tolerates truth-value \emph{gluts}. Another option is to give up the law of excluded middle, so that a formula and its negation may both fail. The resulting negation is then paracomplete, which on the semantics side means that it tolerates truth-value \emph{gaps}. In this paper we are going to study the behaviour of paraconsistent and/or paracomplete logics and see how a mixture of two glut-tolerant and/or gap-tolerant negations can be used to regain the classicality in a natural way.

Our starting point is a family of paraclassical logics  frequently used in the Ghent tradition, namely \(\CLoN, \CLaN\) and \(\CLuN\) \cite{Batens1980ParaconsistentEP,Batens1999embedding}. The logic $\CLoN$ has an ``absolutely free'' negation $\sim_o$, which can be thought of as an uninterpreted operation that allows for both gluts and gaps. The logic \(\CLaN\) has a negation $\sim_a$ which allows for gaps (hence the letter ``a'') and the logic \(\CLuN\) has a negation $\sim_u$  allows for gluts (hence the letter ``u''). These logics have been alternatively used as a basis for non-classical adaptive logics  \cite{Batens1999embedding} and logics of formal inconsistency (LFI) \cite{CarnielliConiglio2016Paraconsistent,CarnielliMarcos2002Taxonomy}. Interestingly, these logics have a very nice  two-valued semantics. This semantics is based on \emph{non-deterministic} matrices in the sense of Avron \cite{Avron2005,Avron2011}. Roughly speaking, in non-deterministic semantics connectives are read as multifunctions, i.e. functions that return a non-empty set of values instead of a single unique value, as it happens in the classical deterministic many-valued semantics.


The heart of the paper is the combination of paraclassical negations. Pairwise combining the one-negation systems yields three two-negations paraclassical logics, $\CLoaN$, $\CLouN$, and $\CLauN$.  In Section \ref{sec:TwoNeg}, we define two-valued non-deterministic semantics for each system and prove  strong soundness and completeness. We focus then on $\CLauN$, which is, in our opinion, the most interesting of the three systems. The
primitive vocabulary of $\CLauN$ contains the gap-tolerant
$\aneg$ alongside the glut-tolerant $\uneg$, which makes
the language considerably more expressive than either one-negation fragment.
In Section \ref{sec:Expr}, we study the expressivity of $\CLauN$ by focusing on the definability of other negation connectives, including classical negation $\cneg$ and the negation $\sim^{au}_o$. The latter can be seen as a representation  of $\sim_o$ within the language of  $\CLauN$. An elegant result connecting the four negations $\aneg,\cneg,\sim_o^{au},\uneg$ is that they form a diamond lattice structure.
Moreover, we
show that it is possible to define in \(\CLauN\) two typical operators of LFI:  a consistency operator and a determinacy operator.
In Section~\ref{sec:ClasRec}, we recover classical propositional logic inside
$\CLauN$ in two ways: locally, by adding classicality assumptions to the
premises, and globally, by a translation that carries the extra information as a parameter directly encoded to the translation by a fresh classical atom, leaving the premises untouched. Finally, in Section \ref{sec:duality} we obtain a duality theorem for the notion of consequence of $\CLauN$.

\section{Technical Preliminaries}\label{sec:intro}

This section is devoted to the technical preliminaries of the paper. In particular, we recall the basics of non-deterministic semantics, and introduce a notion of definability which is specifically tailored to the non-deterministic setting.

On a conceptual level, non-deterministic semantics generalises classical matrix semantics \cite{Wojcicki1984,Wojcicki2013}
by allowing the interpretation of a connective to return a non-empty set of
values, rather than a single value. Accordingly, connectives of the language are interpreted as multi-functions rather than functions.\footnote{If we restrict our attention to finite non-deterministic semantics, this allows for more flexibility and expressive power. Note, however, that when we move to infinite matrices the characterization power of non-deterministic semantics and the usual matrix semantics is the same \cite{Avron2011}.} A valuation may then pick any value
from the set of truth values, and the classical matrices are recovered as the special
case in which every such set is a singleton. Non-deterministic semantics is permissive enough to characterize a wide class of philosophically interesting systems, including modal logics \cite{Coniglio2019,Coniglio2024,Graetz2022,Kearns1981,Omori2020,Pawlowski2023a,Pawlowski2024,Pawlowski2022, ConiglioPawlowskiSkurt2025,Pawlowski2024a}, paraconsistent logics  \cite{Avron2005b,Avron2007d,Avron2009,Omori2019},  logics of informal provability \cite{Pawlowski2018a,Pawlowski2021a,Pawlowski2018,Pawlowski2022}, and even quantum logic \cite{Jorge2020}.  The central notion of the this semantics is the notion of a non-deterministic matrix, or n-matrix for short.

\begin{definition}[N-matrix]\label{def:matrix}
	Given a formal propositional language $\mathcal{L}$, let an n-matrix for $\mathcal{L}$ be a tuple $\mathcal M=\langle V,D,C\rangle$  where:
	\begin{itemize}
		\item \(V\neq \emptyset\) is the set of truth values;
		\item \(\emptyset \neq D\subseteq V\) is the set of designated truth values;
		\item \(C\) is a set of operations \(\overline c:V^n\to 2^V \setminus \{\emptyset\}\), one for each \(n\)-ary connective \(c\) of \(\mathcal L\).
	\end{itemize}
	\noindent An n-matrix is deterministic iff $\overline{c}(x_1,\ldots,x_n)$ is a singleton for all $\overline{c} \in C$ and all $(x_1,\ldots,x_n) \in V^n$.
\end{definition}

\begin{definition}[Valuation]
	Given an n-matrix $\mathcal{M}$ for $\mathcal{L}$, let a non-deterministic valuation in $\mathcal{M}$, or $\mathcal{M}$-valuation, be a function \(v:\Forms_{\mathcal L}\to V\) such that, for every \(n\)-ary connective \(c\) of \(\mathcal L\) and all formulas of $\mathcal L$
	\(A_1,\ldots,A_n \in \Forms_{\mathcal L}\),
	$$
		v(c(A_1,\ldots,A_n)) \in \overline c(v(A_1),\ldots,v(A_n)).
	$$
	\noindent Let $\mathsf{Val}_\mathcal{M}$ be the set of all $\mathcal{M}$-valuations. Given $\{A\},\Gamma \subseteq \Forms_{\mathcal L}$, let  $A$ be a consequence of $\Gamma$ in $\mathcal{M}$, written \(\Gamma\models_{\mathcal M} A\), iff, for every
	\(v \in \mathsf{Val}_\mathcal{M}\), if \(v(B)\in D\) for every
	\(B\in\Gamma\), then \(v(A)\in D\).
\end{definition}

Moving towards definability issues, it often happens in classical many-valued contexts that a few distinct properties closely related with definability are equivalent. For instance, if two formulas have the same truth-tables, then one can substitute one expression with another one without altering the value of the complex statement. However, in a non-deterministic setting we need to be more careful. Due to the fact that connectives of the language are understood as multifunctions, having the same truth table no longer guarantees inter-substitutability.
Hence, we propose the following generalization of the notion of definability.

\begin{definition}[Table definability]\label{def:n-definability}
	Let \(\mathcal M=\langle V,D,C\rangle\) be an n-matrix for
	\(\mathcal L\), and let \(f:V^n\to 2^V \setminus\{\emptyset\}\) be an
	\(n\)-ary operation. A formula \(A(p_1,\ldots,p_n)\) table defines
	\(f\) in \(\mathcal M\) iff, for every
	\((x_1,\ldots,x_n)\in V^n\),
	\[
		\{\,v(A(p_1,\ldots,p_n)) : v\in\Val_{\mathcal M}
		\text{ and }v(p_i)=x_i\text{ for every }1\leq i\leq n\,\}
		=
		f(x_1,\ldots,x_n).
	\]
\end{definition}

Intuitively, table definability generalises the familiar truth-table
criterion for explicit definition: the \emph{definiens} must induce exactly the
table of the operation it defines. The same idea yields a symmetric
relation between formulas.

\begin{definition}[Table equivalence]\label{def:table-equivalence}
	Let \(\mathcal M=\langle V,D,C\rangle\) be an n-matrix for
	\(\mathcal L\). Let \(A(p_1,\ldots,p_n)\) and
	\(B(p_1,\ldots,p_n)\) be formulas whose variables are among
	\(p_1,\ldots,p_n\). We say that \(A(p_1,\ldots,p_n)\) and
	\(B(p_1,\ldots,\) \(p_n)\) are table equivalent in \(\mathcal M\)
	iff there is an \(n\)-ary operation
	\(f:V^n\to 2^V\setminus\{\emptyset\}\) such that \(A(p_1,\ldots,p_n)\) and \(B(p_1,\ldots,p_n)\) both table define $f$.
\end{definition}

Table equivalence is distinct from, but related to, semantic equivalence.

\begin{definition}[Semantic equivalence]\label{def:semantic-equivalence}
	Given an n-matrix \(\mathcal M\) for \(\mathcal L\), formulas
	\(A,B\in\Forms_{\mathcal L}\) are semantically equivalent in
	\(\mathcal M\), written \(A\equiv_{\mathcal M}B\), iff, for every
	\(\mathcal M\)-valuation \(v\), \(v(A)=v(B)\).
\end{definition}

Semantic equivalence always implies table equivalence, once the relevant
variables are fixed. In deterministic matrices, the converse also holds,
and so the two notions coincide. In non-deterministic matrices, however,
table equivalence need not imply semantic equivalence: two formulas may be \emph{table equivalent} without being \emph{semantically equivalent}, i.e. they may
induce the same set-valued table while still receiving different values
under one and the same valuation.

\section{Mixing Negations}\label{sec:TwoNeg}

In this section we introduce the well known one-negation paraclassical logics $\CLoN$, $\CLaN$ and $\CLuN$, as well as the new two-negations paraclassical logics $\CLoaN$, $\CLouN$ and $\CLauN$. Moreover, we show all logics to be sound and complete with respect to suitable non-deterministic matrices.

Axiom systems for the one-negation and two-negations systems can be introduced uniformly by noting that they all contain the axioms and rules of \emph{positive logic}, although in different languages. Then, let $\mathcal{L}_p = \{\land,\lor,\to\}$ and let $\mathcal{L}_x \supseteq \mathcal{L}_p$, for $x \in \{o,a,u,oa,ou,au\}$ be obtained by adding the corresponding negation(s) $\oneg,\aneg,\uneg$.  Furthermore, we will use \(A\leftrightarrow B\) as abbreviations for \((A\to B)\land(B\to A)\). Let $\Forms_{\mathcal{L}_x}$ be the set of formulas generated from a denumerable set \(\Var\) and $\mathcal{L}_x$. Specificially, we use  \(\mathcal{L}_{au}\) for the language of the logic \(\CLauN\), which contains $\sim_a,\sim_u$. Finally, let derivability of $\Delta$ from $\Gamma$ in a Hilbert system for a logic $\mathsf{L}$ ($\Gamma \vdash_{\mathsf{L}} \Delta$) over a language $\mathcal{L}$ be defined as usual.

\begin{definition}[Positive logic]
	Let a Hilbert system for positive logic consist of the following axioms and rule of inference over $\mathcal{L}_p$:
	\begin{itemize}
		\item [\textbf{(P1)}] \(A\to(B\to A)\)
		\item [\textbf{(P2)}] \((A\to(B\to C))\to((A\to B)\to(A\to C))\)
		\item [\textbf{(P3)}] \(((A\to B)\to A)\to A\)
		\item [\textbf{(P4)}] \((A\land B)\to A\)
		\item [\textbf{(P5)}] \((A\land B)\to B\)
		\item [\textbf{(P6)}] \((C\to A)\to((C\to B)\to(C\to(A\land B)))\)
		\item [\textbf{(P7)}] \(A\to(A\lor B)\)
		\item [\textbf{(P8)}] \(B\to(A\lor B)\)
		\item [\textbf{(P9)}] \((A\to C)\to((B\to C)\to((A\lor B)\to C))\)
		\item [\textbf{(P10)}] \( \dfrac{A \quad A \to B}{B} \).
	\end{itemize}
\end{definition}

From axioms and rules \textbf{(P1)}-\textbf{(P10)}, one-negation logics are defined as follows. $\CLoN$ is defined by extending the language with a paracomplete and paraconsistent negation $\sim_o$; $\CLaN$ is defined by extending the language with a gap-tolerant only negation $\aneg$, i.e. $\aneg$ is paracomplete but not paraconsistent; and $\CLuN$ is defined by extending the language with a glut-tolerant only negation $\uneg$, i.e. it is paraconsistent but not paracomplete. Two-negations logics $\CLoaN$, $\CLouN$ and $\CLauN$ are defined by mixing and matching one-negation logics.

\begin{definition}[Paraclassical logics]
	Let the following Hilbert system for paraclassical logics be defined over $\mathcal{L}_x$ and $\mathcal{L}_{yz}$, for $x \in \{\mathsf{o},\mathsf{a},\mathsf{u}\}$ and $yz \in \{\mathsf{oa},\mathsf{ou},\mathsf{au}\}$:
	$$
		\mathsf{CL}x\mathsf{N} := \textbf{(P1)}-\textbf{(P10)} + \mathsf{x} \qquad \mathsf{CL}yz\mathsf{N} := \textbf{(P1)}-\textbf{(P10)} + \mathsf{y} + \mathsf{z}
	$$
	\noindent where $\mathsf{o}$ is the empty string, $\mathsf{a} := A \to ( \aneg A \to B)$ and $\mathsf{u} := A \lor \uneg A$.
\end{definition}

With an abuse of notation, we use in what follows $\mathsf{L}$ and $\mathcal{L}$ as meta-variables over one-negation and two-negations paraclassical logics and their corresponding sets of connectives. Non-deterministic matrices for all $\mathsf{L}$ are obtained simply by specifying the truth tables for the operations associated with each connective. As expected, $\land,\lor,\to$ are associated with deterministic operations, while $\oneg,\aneg,\uneg$ are associated with genuine non-deterministic operations.

\begin{definition}[N-matrices]\label{def:n--matrix}
	Let an n-matrix for $\mathsf{L}$ be a tuple $\mathcal{M}_{\mathsf{L}} = \langle \{1,0\},\{1\},C \rangle$, where the truth tables for all operations $\overline{c} \in C$, one for each $c \in \mathcal{L}$, are specified in Table \ref{tab:1}.
\end{definition}


\begin{table}[t]
	\begin{framed}
		\[
			\begin{array}{c|cc}
				\overline{\land} & 1     & 0     \\ \hline
				1                & \{1\} & \{0\} \\
				0                & \{0\} & \{0\}
			\end{array}
			\ \
			\begin{array}{c|cc}
				\overline{\lor} & 1     & 0     \\ \hline
				1               & \{1\} & \{1\} \\
				0               & \{1\} & \{0\}
			\end{array}
			\ \
			\begin{array}{c|cc}
				\overline{\to} & 1     & 0     \\ \hline
				1              & \{1\} & \{0\} \\
				0              & \{1\} & \{1\}
			\end{array}
			\ \
			\begin{array}{c|c|c|c}
				  & \overline{\oneg} & \overline{\aneg} & \overline{\uneg} \\ \hline
				1 & \{1,0\}          & \{0\}            & \{1,0\}          \\
				0 & \{1,0\}          & \{1,0\}          & \{1\}
			\end{array}
		\]
	\end{framed}
	\caption{Non-deterministic tables for $\land,\lor,\to,\oneg,\aneg,\uneg$.}
	\label{tab:1}
\end{table}

Soundness and completeness for each $\mathsf{L}$ with respect to its corresponding matrix $\mathcal{M}_{\mathsf{L}}$ is obtained as follows, via the usual methods of \cite{Avron2005,Batens1999embedding,Omori2021}.

\begin{theorem}[Soundness and completeness]\label{thm:soundness}\label{thm:completeness}\label{cor:sc}
	For \(\mathsf{L}\in\{\CLoaN,\CLouN,\CLauN\}\)
	\[\Gamma\vdash_\mathsf{L} A \ \text{ iff } \ \Gamma\vDash_{\mathcal{M}_\mathsf{L}} A.\]
\end{theorem}

\begin{proof}
	The left to right direction is established by a routine induction on the length of $\mathsf{L}$-derivations.
	The right to left direction is established as follows. Assuming $\Gamma \not\vdash_{\mathsf{L}} A$, we show that there is an $\mathcal{M}_\mathsf{L}$-valuation such that $\Gamma \not\models_{\mathsf{L}} A$. To this aim, note that $\Gamma \not\vdash_{\mathsf{L}} A$ implies by the well known pair extension theorem that there is a prime $\mathsf{L}$-theory $\Sigma$ such that \(\Gamma\subseteq\Sigma\) and \(A\notin\Sigma\). Then, define the desired valuation so that
	$$
		v_\Sigma(B) := \begin{cases}
			1 & \text{if } \Sigma \vdash_{\mathsf{L}} B      \\
			0 & \text{if } \Sigma \not\vdash_{\mathsf{L}} B.
		\end{cases}
	$$
	\noindent To verify that $v_\Sigma$ is indeed an $\mathcal{M}_\mathsf{L}$-valuation, we show that $v(c(A_1,\ldots,A_n)) \in \overline c(v(A_1),\ldots,v(A_n))$ by induction on the complexity of $B$, in the language of $\mathsf{L}$. The base case $B \in \mathsf{Var}$ and the induction step cases involving $c \in \{\land,\lor,\to,\oneg\}$ are immediate.
	If $B= \uneg C$ (hence $\mathsf{L} \in \{\CLuN,\CLouN,\CLauN\}$), then we need to show that  $v_{\Sigma}(\uneg C) \in \overline{\uneg} v_\Sigma(C)$. If $v_\Sigma(C)= 0$, then by induction hypothesis $C \notin \Sigma$. By regularity of $\Sigma$ and $\vdash_{\mathsf{L}} C \lor \uneg C$, we infer that $C \lor \uneg C \in \Sigma$, whence $\uneg C \in \Sigma$ by primeness of $\Sigma$.  We conclude that $v_\Sigma (\uneg C) = 1 \in \overline{\uneg} 0 = \{1\}$. If $v_\Sigma(C)= 1$ the result is immediate.
	If $B= \aneg C$ (hence $\mathsf{L} \in \{\CLaN,\CLoaN,\CLauN\}$), then we need to show that  $v_{\Sigma}(\aneg C) \in \overline{\aneg} v_\Sigma(C)$. If $v_\Sigma(C)= 1$, assume by contradiction that $v_{\Sigma}(\aneg C) \notin \overline{\aneg} 1$. Hence, $v_{\Sigma}(\aneg C) = 1$, which implies by $v_\Sigma(C)= 1$ and the induction hypothesis that both $C \in \Sigma$ $\aneg C \in \Sigma$. We conclude by $\vdash_{\mathsf{L}} C \to ( \aneg C \to A)$ that $A \in \Sigma$, which is a contradiction. If $v_\Sigma(C)= 0$ the result is immediate.
	We conclude that $v_\Sigma$ is an $\mathcal{M}_\mathsf{L}$-valuation such that $v_{\Sigma}(G) = 1$ for all $G \in \Gamma \subseteq \Sigma$ and  $v_{\Sigma}(A) = 0$ (since $A \notin \Sigma$), whence $\Gamma \not\models_{\mathcal{M}_\mathsf{L}} A$.
\end{proof}

In the remainder of the article, we focus on the system $\CLauN$ over $\CLoaN$ and $\CLouN$. The reason is that $\CLauN$ is a natural environment to establish connections with logics of formal inconsistency and classical logic, as well as to obtain definability and duality results. We return to systems $\CLoaN$ and $\CLouN$ in Section \ref{sec:Concl}.

\section{Expressivity}\label{sec:Expr}

In this section, we show some semantic definability results for our preferred two-negations paraclassical logic, $\CLauN$, partly in relation with so-called logics of formal inconsistency \cite{Avron2009,Carnielli2007,Costa1977}.

Our first result concerns table definability in the system $\CLauN$. Specifically, we will take a look at missing negations: the non-deterministic negation $\oneg$ of logic $\CLoN$, and the classical negation  $\cneg$, whose deterministic table is obtained as usual, by putting $\overline{\cneg} x = \{1 - x\}$.

\begin{theorem}[Table definability]
	\label{thm:expressive-comparison}
	The following facts hold in $\CLauN$:
	\begin{enumerate}[label=\textup{(\roman*)},leftmargin=2em]
		\item $p\to\aneg p$ table defines $\overline{\cneg}$ in $\mathcal{M}_{\CLauN}$.

		\item \(\aneg\aneg p\) table defines  $\overline{\oneg}$  in $\mathcal{M}_{\CLauN}$.
		\item \(\uneg p \land (p \lor \aneg p)\) table defines   $\overline{\oneg}$  in $\mathcal{M}_{\CLauN}$.
	\end{enumerate}
\end{theorem}
\begin{proof}
	Let \(v\) be an arbitrary \(\mathcal M_{\CLauN}\)-valuation. For \textup{(i)}, assume that \(v(p)=1\), then \(v(\aneg p)=0 \) and   \(v( p\to\aneg p)=0\) as it should. If \(v(p)=0\), then the antecedent of the implication is also \(0\) and so the value of the implication is \(1\). For \textup{(ii)}, observe that the value of \(\overline{\oneg}\) is always in \(\{0,1\}\). If  \(v(p)=1\), then the value of \(\aneg\aneg p\) is in \(\{0,1\}\) and the same holds if \(v(p)=0\). For \textup{(iii)}, if \(v(p)=0\), then \(v(\uneg p)=1\), which means that the value of \(\uneg p \land (p \lor \aneg p)\) depends on the value of $\aneg p$ which in this particular case can be either \(0\) or \(1\). If  \(v(p)=1\), then the value of  \(\uneg p \land (p \lor \aneg p)\) solely depends on the value of $\uneg p$ which, again, can be \(0\) or \(1\).
\end{proof}

Some comments regarding the above theorem are in order. First, we stress again that two table definitions of the same operation need not be semantically equivalent.\footnote{This can now be seen by checking that  \(\aneg\aneg p\) (definition in (\textup{(ii)} ) is not semantically equivalent to \(\uneg p \land (p \lor \aneg p)\) (definition in \textup{(iii)}). Take a valuation $v$ such that $v(p)=0$ and let $v(\aneg p)=0$.  Furthermore, we may choose that $v(\aneg \aneg p)=1$. On the other hand, $v(\uneg p)=1$, so $(v(p\lor \aneg p))=0$, hence $v(\uneg p \et (p \lor \aneg p))=0$ giving the lack of the semantic equivalence.}
Second, recall that table definability is the same as semantic equivalence in case of deterministic tables, so in effect $A \to \aneg A$ is not only table equivalent but semantically equivalent to $\cneg A$.
Finally, there are reasons to prefer one particular definition of $\oneg$ over the other. In our case, it is preferable to syntactically abbreviate $\sim_o^{au} A := \uneg A \land (A \lor \aneg A)$, because it will be seen that $\uneg A \land (A \lor \aneg A)$ is semantically equivalent to $\aneg A \lor (A \land \uneg A)$. The latter fact allows one to connect, in terms of semantic equivalence and not just table equivalence, classical negation $\cneg$ and $\sim_o^{au} A$ on the one side,  with the primitive $\CLauN$ negations $\aneg$ and $\uneg$ on the other side. Finally, this turns out essential to show the main result of Section \ref{sec:ClasRec}, i.e. a recapture of classical propositional logic within $\CLauN$.

\begin{theorem}[Semantic equivalence]
	\label{thm:boundary-decomposition}
	The following facts hold for $\CLauN$:
	\begin{enumerate}[label=\textup{(\roman*)},leftmargin=2em]
		\item \(\sim_o^{au} A\equiv
		      \aneg A\lor(A\land\uneg A)\).

		\item \(\aneg A\equiv\cneg A\land \sim_o^{au} A\).

		\item \(\uneg A\equiv\cneg A\lor \sim_o^{au} A\).
	\end{enumerate}
\end{theorem}
\begin{proof}
	Let \(v\) be an arbitrary \(\mathcal M_{\CLauN}\)-valuation. For \textup{(i)}, if \(v(A)=1\), then $v(\aneg A)=0$ and \(v(A \lor \aneg A)=1\). The former implies that \(v(\aneg A\lor(A\land\uneg A))=v(\uneg A)\), while the latter implies that \(v(\sim_o^{au} A) = v(\uneg A)\). Hence $v(\sim_o^{au} A) = v(\aneg A\lor(A\land\uneg A))$. If \(v(A)=0\), then
	\(v(\uneg A)=1\) and $v(A \land \uneg A)=0$. The former implies that \(v(\sim_o^{au} A) = v(\aneg A)\), while the latter implies that  \(v(\aneg A\lor(A\land\uneg A))=v(\aneg A)\). Hence
	\(v(\sim_o^{au} A) = v(\aneg A\lor(A\land\uneg A))\).
	For \textup{(ii)}, if \(v(A)=1\), then \(v(\cneg A)=0=v(\aneg A)\),
	so \(v(\cneg A\land \sim_o^{au} A)=v(\aneg A)\). If \(v(A)=0\), then
	\(v(\cneg A)=1\) and \(v(\sim_o^{au} A)=v(\aneg A)\), so
	\(v(\cneg A\land \sim_o^{au} A)=v(\aneg A)\).
	For \textup{(iii)}, if \(v(A)=1\), then \(v(\cneg A)=0\) and
	\(v(\sim_o^{au} A)=v(\uneg A)\), so \(v(\cneg A\lor \sim_o^{au} A)=v(\uneg A)\). If
	\(v(A)=0\), then \(v(\cneg A)=1=v(\uneg A)\), so
	\(v(\cneg A\lor \sim_o^{au} A)=v(\uneg A)\).
\end{proof}

An interesting consequence of the above theorem is captured by the following proposition.

\begin{proposition}[Negations]\label{thm:diamond} The following facts hold for $\CLauN$:: \begin{enumerate}[label=\textup{(\roman*)},leftmargin=2em]
\item \(\aneg A\vDash_{\CLauN}\cneg A \vDash_{\CLauN}\uneg A\). 
\item \(\aneg A\vDash_{\CLauN}\sim_o^{au}A \vDash_{\CLauN}\uneg A\).
\item  There exist $A \in \Forms_{\mathcal L_{au}}$ such that \(\cneg A\nvDash_{\CLauN}\sim_o^{au}A\). 
\item There exist $A \in \Forms_{\mathcal L_{au}}$ such that \(\sim_o^{au}A\nvDash_{\CLauN}\cneg A\). \end{enumerate} 
\end{proposition}

\begin{proof}
	\textup{(i)} and
	\textup{(ii)} are immediate consequences of Theorem~\ref{thm:boundary-decomposition}. For (iii), take a
	valuation such that \(v(p)=0\) and \(v(\aneg p)=0\). Then
	\(v(\uneg p)=1\), and so \(v(\sim_o^{au}p)=
	v(\uneg p\land(p\lor\aneg p))=0\), while \(v(\cneg p)=1\).  Hence, \(\cneg p\nvDash_{\mathcal{M}_{\CLauN}}\sim_o^{au}p\). Conversely, take a
	valuation such that \(v(p)=1\) and \(v(\uneg p)=1\). Then
	\(v(\aneg p)=0\), and so \(v(\sim_o^{au}p)=1\), while
	\(v(\cneg p)=0\). Hence, \(\sim_o^{au}p\nvDash_{\mathcal{M}_{\CLauN}}\cneg p\).
\end{proof}

\begin{remark}[Negation diamond]\label{rmk:Diamond}
	The preceding result gives an elegant representation of the relationships between the four
	negations $\aneg$, $\uneg$, $\sim_o^{au}$ and $\cneg$. These relationships can be depicted by what we may dub the negation diamond, where lines indicate derivability:
	\[
		\begin{tikzpicture}[baseline=(current bounding box.center),node distance=1.25cm]
			\node (top) {\(\uneg A\)};
			\node (left) [below left of=top,xshift=-.4cm] {\(\cneg A\)};
			\node (right) [below right of=top,xshift=.4cm] {\(\sim_o^{au} A\)};
			\node (bottom) [below right of=left,xshift=.4cm] {\(\aneg A\)};

			\draw[->] (bottom) -- (left);
			\draw[->] (bottom) -- (right);
			\draw[->] (left) -- (top);
			\draw[->] (right) -- (top);
		\end{tikzpicture}
	\]
	\noindent That is, with respect to a given formula $A$, paracomplete negation implies both classical and non-deterministic negation, which are incomparable, and these in turn imply paraconsistent negation.
\end{remark}

We close this section with some remarks which relate \(\CLauN\) with logics of formal inconsistency LFI \cite{Avron2009,Carnielli2007,Costa1977}. A defining
feature of LFI is that they are paraconsistent, i.e.
\(A,\neg A\nvdash B\), but explosion can be locally recovered by
assuming that \(A\) behaves consistently. This is usually done by introducing a primitive connective to the object language. We pursue a related but slightly different strategy. Although \(\CLauN\) does not
include a consistency connective as primitive, we can syntactically define natural candidates that could serve as consistency operators. We propose three definitions, relying on the combinations of paraclassical negations of the object language:
$$
	\circ_u A\defeq \cneg(A\land\uneg A) \qquad  \delta_a A\defeq A\lor\aneg A \qquad \mathsf{Cl}A\defeq \circ_u A\land\delta_a A.
$$
The idea of the consistency  operator  \(\circ_u A\defeq \cneg(A\land\uneg A)\) is to rule out the case in which both \(A\) and \(\uneg A\) are true. The idea of the determinacy operator \(\delta_a A\defeq A\lor\aneg A\) is to rule out the case in which both \(A\) and \(\aneg A\) fail. Finally, the idea of the classicality operator \( \mathsf{Cl}A\defeq \circ_u A\land\delta_a A\) is to ensure that the formula behaves consistently and in a determined way.
In other words, these three operators isolate the two local forms of
non-classicality available in the system. The failure of \(\circ_u A\)
is a local glut, while the failure of \(\delta_a A\) is a local gap. We can make these propositions more precise in the form of the following proposition.

\begin{proposition}[Classicality locally recaptured]\label{prop:local} The following facts hold for any \(A,B\in\Forms_{\mathcal L_{au}}\): \begin{enumerate}[label=\textup{(\roman*)},leftmargin=2em] \item \(\circ_u A,A,\uneg A \vDash_{\CLauN} B\). \item \(\delta_a A,A\to B,\aneg A\to B \vDash_{\CLauN} B\). \item \(\vDash_{\CLauN} \circ_u A \leftrightarrow (\uneg A \leftrightarrow \cneg A)\). \item \(\vDash_{\CLauN} \delta_a A \leftrightarrow (\aneg A \leftrightarrow \cneg A)\). \item \(\vDash_{\CLauN} \mathsf{Cl}A \leftrightarrow (\sim_o^{au} A \leftrightarrow \cneg A)\). \end{enumerate} \end{proposition} \begin{proof} Let \(v\) be an arbitrary \(\mathcal M_{\CLauN}\)-valuation. For \textup{(i)}, it suffices to show that if \(v(A)=1\) and \(v(\uneg A)=1\), then \(v(\circ_u A)=0\). Indeed, \(v(A)=1\) and \(v(\uneg A)=1\) imply \(v(A\land\uneg A)=1\), so \(v(\circ_u A)=v(\cneg(A\land\uneg A))=0\). For \textup{(ii)}, suppose by contradiction that \(v(\delta_a A)=1\), \(v(A\to B)=1\), \(v(\aneg A\to B)=1\), and \(v(B)=0\). From \(v(A\to B)=1\) and \(v(B)=0\), we get \(v(A)=0\). From \(v(\aneg A\to B)=1\) and \(v(B)=0\), we get \(v(\aneg A)=0\). But then \(v(\delta_a A)=v(A\lor\aneg A)=0\), contradicting \(v(\delta_a A)=1\). For \textup{(iii)}, if \(v(A)=1\), then \(v(\cneg A)=0\), and so \(v(\circ_u A)=1\) iff \(v(\uneg A)=0\). If \(v(A)=0\), then \(v(\cneg A)=1\), \(v(\uneg A)=1\), and \(v(\circ_u A)=1\). Thus \(v(\circ_u A)=1\) iff \(v(\uneg A)=v(\cneg A)\). Since \(\leftrightarrow\) abbreviates \((\to)\) in both directions, \((\uneg A\leftrightarrow\cneg A)\) has value \(1\) exactly when \(v(\uneg A)=v(\cneg A)\). Hence \(\vDash_{\CLauN} \circ_u A \leftrightarrow (\uneg A\leftrightarrow\cneg A)\). For \textup{(iv)}, if \(v(A)=1\), then \(v(\cneg A)=0\), \(v(\aneg A)=0\), and \(v(\delta_a A)=1\). If \(v(A)=0\), then \(v(\cneg A)=1\), and \(v(\delta_a A)=1\) iff \(v(\aneg A)=1\). Thus \(v(\delta_a A)=1\) iff \(v(\aneg A)=v(\cneg A)\). Hence \(\vDash_{\CLauN} \delta_a A \leftrightarrow (\aneg A\leftrightarrow\cneg A)\). For \textup{(v)}, by \textup{(iii)} and \textup{(iv)}, we have \(v(\mathsf{Cl}A)=1\) iff \(v(\cneg A)=v(\aneg A)=v(\uneg A)\). By Theorem~\ref{thm:boundary-decomposition}, this is equivalent to \(v(\sim_o^{au}A)=v(\cneg A)\). Hence \(\vDash_{\CLauN} \mathsf{Cl}A \leftrightarrow (\sim_o^{au}A\leftrightarrow\cneg A)\). \end{proof}

\section{Full Classical Recapture}\label{sec:ClasRec}

In this section, we show two ways in which it is possible to recapture classical propositional logic $\mathsf{CL}$ within $\CLauN$. Our first result exploits the close relationship between $\CLauN$ and logics of formal inconsistency \cite{Carnielli2007} to show that it is possible to recover classical reasoning in $\CLauN$. We do so by assuming that the formulas locally involved  in an argument are classical.  To this aim, let $\mathrm{Arg}_x(S) :=\{B:{\sim_x}B\text{ occurs in some }C\in S \subseteq\Forms_{\mathcal{L}_{au}}\}$, for every $x \in \{a,u\}$ and $S \subseteq \Forms_{\mathcal{L}_{au}}$ . Informally, this set consists of all formulas which occur within the scope of a negation \(\sim_x\) somewhere in the formulas belonging to \(S\). Based on this, we define the set of classical formulas $\mathbb{CLS}(S) \subseteq\Forms_{\mathcal{L}_{au}}$:
$$
	\mathbb{CLS}(S) :=
	\{\circ_u B:B\in\mathrm{Arg}_u(S)\}
	\cup
	\{\delta_a B:B\in\mathrm{Arg}_a(S)\}.
$$
\noindent Hence, $\mathbb{CLS}(S)$ contains exactly those formulas which make the formulas in $S$ behave classically, i.e. $\uneg$-consistently and $\aneg$-completely.

\begin{definition}[First translation] Let \(\mathcal L_{\mathrm{cl}}\) be the classical language over the same propositional variables, and let \(\mathbf{Tr}:\Forms_{\mathcal L_{au}}\to \Forms_{\mathcal L_{\mathrm{cl}}}\) be defined recursively as follows:
	\[ \mathbf{Tr}(A)= \begin{cases} p, & \text{if } A=p,\\[1mm] \mathbf{Tr}(B)\star \mathbf{Tr}(C), & \text{if } A=B\star C \text{ and } \star\in\{\land,\lor,\to\},\\[1mm] \cneg \mathbf{Tr}(B), & \text{if } A=\uneg B \text{ or } A=\aneg B \end{cases} \] For \(\Gamma\subseteq\Forms_{\mathcal L_{au}}\), let \(\mathbf{Tr}[\Gamma]=\{\mathbf{Tr}(G):G\in\Gamma\}\).
\end{definition}

This is a straightforward translation that collapses non classical negations into the classical negation. By making sure that all formulas involved in an argument behave classically,  i.e.  belong to \(\mathbb{CLS}(S)\),  we can prove the following theorem.

\begin{proposition}[First classical recapture] \label{prop:finite-recapture} For every finite \(S\subseteq\Forms_{\mathcal L_{au}}\) and every \(\{A\}\cup\Gamma\subseteq S\), \[\mathbb{CLS}(S),\Gamma\vDash_{\CLauN} A \quad\text{iff}\quad \mathbf{Tr}[\Gamma]\vDash_{\CL}\mathbf{Tr}(A). \]
\end{proposition}

\begin{proof} First observe that, for every \(\mathcal M_{\CLauN}\)-valuation \(v\) such that \(v(D)=1\) for all \(D\in\mathbb{CLS}(S)\), if \(e_v\) is the classical valuation defined by \(e_v(p)=v(p)\), then \(v(C)=e_v(\mathbf{Tr}(C))\) for every subformula \(C\) of a formula in \(S\). This is proved by induction on \(C\). The atomic case and the cases for \(\land,\lor,\to\) are immediate. If \(C=\uneg B\), then \(\circ_u B\in\mathbb{CLS}(S)\), so \(v(\uneg B)=v(\cneg B)\) by Proposition~\ref{prop:local}. By the induction hypothesis, \(v(B)=e_v(\mathbf{Tr}(B))\), whence \(v(\uneg B)=e_v(\cneg\mathbf{Tr}(B))=e_v(\mathbf{Tr}(\uneg B))\). The case \(C=\aneg B\) is analogous, using \(\delta_a B\in\mathbb{CLS}(S)\). Suppose first that \(\mathbf{Tr}[\Gamma]\vDash_{\CL}\mathbf{Tr}(A)\). Let \(v\) be an \(\mathcal M_{\CLauN}\)-valuation such that \(v(D)=1\) for all \(D\in\mathbb{CLS}(S)\) and \(v(G)=1\) for all \(G\in\Gamma\). By the observation above, \(e_v(\mathbf{Tr}(G))=1\) for all \(G\in\Gamma\). Hence \(e_v(\mathbf{Tr}(A))=1\), and so \(v(A)=1\). Therefore \(\mathbb{CLS}(S),\Gamma\vDash_{\CLauN} A\). Conversely, suppose that \(\mathbf{Tr}[\Gamma]\nvDash_{\CL}\mathbf{Tr}(A)\). Let \(e\) be a classical valuation such that \(e(\mathbf{Tr}(G))=1\) for all \(G\in\Gamma\), while \(e(\mathbf{Tr}(A))=0\). Define an \(\mathcal M_{\CLauN}\)-valuation \(v_e\) by putting \(v_e(p)=e(p)\), interpreting \(\land,\lor,\to\) classically, and choosing \(v_e(\uneg B)=v_e(\aneg B)=1-v_e(B)\) for every formula \(B\). This is allowed by the tables for \(\uneg\) and \(\aneg\). By induction, \(v_e(C)=e(\mathbf{Tr}(C))\) for every \(C\in\Forms_{\mathcal L_{au}}\). Thus \(v_e(G)=1\) for all \(G\in\Gamma\) and \(v_e(A)=0\). Moreover, every formula in \(\mathbb{CLS}(S)\) has value \(1\) under \(v_e\): if \(D=\circ_u B\), then \(v_e(B\land\uneg B)=0\), and if \(D=\delta_a B\), then \(v_e(B\lor\aneg B)=1\). Hence \(\mathbb{CLS}(S),\Gamma\nvDash_{\CLauN} A\).
\end{proof}

Our second result gives another translation of \(\CLauN\) into \(\mathsf{CL}\). It lifts the previous recapture result by keeping the formulas involved in an argument unchanged and building the needed additional information directly into the translation. To this end, we work with an expanded classical language.\footnote{As noted by a careful reviewer, compared to the first translation, here we are here, strictly speaking, translating $\CLauN$ not into $\mathsf{CL}$ itself, but into a theory extending $\mathsf{CL}$ with propositional atoms.}  Let \(\mathcal L_{cl}\) be the classical propositional language with connectives \(\land,\lor,\to,\neg\) and atoms \(\Var\cup\{\mathfrak o_A:A\in\Forms_{\mathcal L_{au}}\}\), where \(\Var\) is the set of atoms of the original language and each \(\mathfrak o_A\) is a fresh atom. The idea here is that these atoms are used to express the extra information that is associated with the formula  \(A\). In other words, the idea is that \(\mathfrak{o}_A\) records the extra truth-value choice associated with the negations of \(A\): when \(A\) is true, it fixes the possible value of \(\uneg A\), and when \(A\) is false, it fixes the possible value of \(\aneg A\).

\begin{definition}[Second translation] Let \(\mathbf{Tr}_2:\Forms_{\mathcal L_{au}}\to\Forms_{\mathcal L_{cl}}\) be the translation defined recursively by: \[ \mathbf{Tr}_2(A)= \begin{cases} p, & \text{if } A=p,\\[1mm] \mathbf{Tr}_2(B)\star\mathbf{Tr}_2(C), & \text{if } A=B\star C \text{ and } \star\in\{\land,\lor,\to\},\\[1mm] \neg\mathbf{Tr}_2(B)\lor\mathfrak o_B, & \text{if } A=\uneg B,\\[1mm] \neg\mathbf{Tr}_2(B)\land\mathfrak o_B, & \text{if } A=\aneg B. \end{cases} \] For \(\Gamma\subseteq\Forms_{\mathcal L_{au}}\), let \(\mathbf{Tr}_2[\Gamma]=\{\mathbf{Tr}_2(G):G\in\Gamma\}\). \end{definition}

\begin{theorem}[Second classical recapture]\label{thm:troupe} For every \(\Gamma\cup\{A\}\subseteq\Forms_{\mathcal L_{au}}\), \[ \Gamma\vDash_{\CLauN} A \quad\text{iff}\quad \mathbf{Tr}_2[\Gamma]\vDash_{\CL}\mathbf{Tr}_2(A). \] \end{theorem}

\begin{proof} First let \(v\) be an \(\mathcal M_{\CLauN}\)-valuation. Define a classical valuation \(e_v\) on \(\Forms_{\mathcal L_{cl}}\) by \(e_v(p)=v(p)\) and \(e_v(\mathfrak o_B)=v(\sim_o^{au}B)\). We show by induction on \(B\) that \(e_v(\mathbf{Tr}_2(B))=v(B)\). The atomic case and the cases for \(\land,\lor,\to\) are immediate. If \(B=\uneg C\), then, by Theorem~\ref{thm:boundary-decomposition}, \(v(\uneg C)=v(\cneg C\lor\sim_o^{au}C)\). By the induction hypothesis, this is exactly \(e_v(\cneg\mathbf{Tr}_2(C)\lor\mathfrak o_C)\), hence \(e_v(\mathbf{Tr}_2(\uneg C))=v(\uneg C)\). The case \(B=\aneg C\) is analogous, using \(v(\aneg C)=v(\cneg C\land\sim_o^{au}C)\). Conversely, let \(e\) be a classical valuation on \(\Forms_{\mathcal L_{cl}}\), and define \(v_e(B)=e(\mathbf{Tr}_2(B))\). We show that \(v_e\) is an \(\mathcal M_{\CLauN}\)-valuation. The cases for \(\land,\lor,\to\) are immediate. If \(v_e(B)=0\), then \(v_e(\uneg B)=e(\neg\mathbf{Tr}_2(B)\lor\mathfrak o_B)=1\), as required by the table for \(\uneg\). If \(v_e(B)=1\), then \(v_e(\uneg B)\) may be either \(0\) or \(1\), again as required. Similarly, if \(v_e(B)=1\), then \(v_e(\aneg B)=e(\neg\mathbf{Tr}_2(B)\land\mathfrak o_B)=0\), while if \(v_e(B)=0\), either value is allowed. Thus \(v_e\) is an \(\mathcal M_{\CLauN}\)-valuation. The equivalence of consequence now follows by the two constructions. A countermodel to \(\Gamma\vDash_{\CLauN}A\) gives, by the first construction, a classical countermodel to \(\mathbf{Tr}_2[\Gamma]\vDash_{\CL}\mathbf{Tr}_2(A)\). Conversely, a classical countermodel to \(\mathbf{Tr}_2[\Gamma]\vDash_{\CL}\mathbf{Tr}_2(A)\) gives, by the second construction, a countermodel to \(\Gamma\vDash_{\CLauN}A\). \end{proof}

\section{Duality}\label{sec:duality} In this section, we show an internal duality theorem for the semantic consequence relation determined by \(\mathcal{M}_{\CLauN}\). A closer inspection of the truth-tables of the glut-tolerant negation \(\uneg\) and the gap-tolerant negation \(\aneg\) suggests a duality between them. This duality has already been used, in a related form, to connect the consequence relations of \(\CLaN\) and \(\CLuN\) \cite{Batens1999embedding} in the context of the interpolation theorem. To the best of our knowledge, however, it has not been formulated as an internal duality theorem for \(\CLauN\).

In order to properly state the result in full generality, we need to work with multiple conclusion consequence, according to which sets of premises are read conjunctively and sets of conclusions are read disjunctively. We restrict our attention to finite sets.

\begin{definition}[Multiple-conclusion consequence]\label{def:mc-consequence} For \(\Gamma,\Pi\subseteq\Forms_{\mathcal{L}_{au}}\), write \(\Gamma\Vdash_{\CLauN}\Pi\) iff there is no \(\mathcal{M}_{\CLauN}\)-valuation \(v\) such that \(v(G)=1\) for every \(G\in\Gamma\) and \(v(P)=0\) for every \(P\in\Pi\).
\end{definition}

Our duality result establishes that $\Gamma\Vdash_{\CLauN}\Pi
	\quad\text{iff}\quad
	\Pi^d\Vdash_{\CLauN}\Gamma^d$, where  $
	X^d=\{A^d:A\in X\}$ is the dual of a set of formulas $X\subseteq\Forms_{\mathcal L_{au}}$.
Intuitively, the role of the dual translation $d: A \mapsto A^d$ is to ensure that for every valuation under which $A$ is true (false), there is a valuation  for which $A^d$ is false (true). Therefore, the clauses of the dual translation reflect the falsity conditions of the connectives. For example, a conjunction is false iff either the first or the second conjunct is false, which explains why conjunction and disjunction are exchanged by the dual translation. However, the situation is slightly different for implication, since it is false exactly when the antecedent is true and the consequent is false. In our setting, neither $\aneg$ nor $\uneg$ reverses truth values deterministically and so we need to construct a formula that forces the antecedent to be true and the consequent to be false. The consequent is not problematic since its dual suffices. For the antecedent, we need a formula that corresponds its truth. We use \(B^d\to\aneg B^d\), since it clearly does the job.\footnote{We note that for $(B \to C)^d$ we use the
	expanded form $(B^d \to \aneg B^d)$ rather than a primitive \(\cneg B^d\). This is because treating
	\(\cneg\) as primitive would make distinct formulas collapse under the
	translation, for example \(p\to q\) and \(\cneg p\lor q\). Another option would be to extend the language with a connective for co-implication. This, however lies beyond the scope of the present paper.}


\begin{definition}[Dual translation]\label{def:dual-translation}
	For every \(A\in\Forms_{\mathcal L_{au}}\), define \(A^d\) recursively by
	\[
		A^d =
		\begin{cases}
			p,                             & \text{if } A=p,        \\[1mm]
			B^d\land C^d,                  & \text{if } A=B\lor C,  \\[1mm]
			B^d\lor C^d,                   & \text{if } A=B\land C, \\[1mm]
			(B^d \to \aneg B^d) \land C^d, & \text{if } A=B\to C,   \\[1mm]
			\uneg B^d,                     & \text{if } A=\aneg B,  \\[1mm]
			\aneg B^d,                     & \text{if } A=\uneg B.
		\end{cases}
	\]
\end{definition}


\begin{lemma}\label{lem:dual-injective}
	The map \(A\mapsto A^d\) is injective and no formula of the form
	\(A^d\) has \(\to\) as its outermost connective.
\end{lemma}
\begin{proof}
	We first show that no formula of the form \(A^d\) has \(\to\) as its
	outermost connective. This is proved by induction on \(A\). If \(A\) is
	atomic, then \(A^d=A\). If \(A\) is \(B\lor C\), \(B\land C\),
	\(B\to C\), \(\aneg B\), or \(\uneg B\), then \(A^d\) has, respectively,
	outermost connective \(\land\), \(\lor\), \(\land\), \(\uneg\), or
	\(\aneg\). Hence the outermost connective of \(A^d\), when there is one,
	is never \(\to\). We now prove injectivity. Suppose that \(A^d=B^d\). We argue by induction on the joint complexity of \(A\) and \(B\). If \(A\) is atomic, then
	\(A^d\) is atomic. Since the only translations that are atomic are
	translations of atoms, \(B\) must be the same atom. Thus \(A=B\). If \(A=A_0\land A_1\), then \(A^d=A_0^d\lor A_1^d\). Hence \(B^d\) also
	has outermost connective \(\lor\), so \(B\) must be of the form
	\(B_0\land B_1\). By syntactic equality, \(A_0^d=B_0^d\) and
	\(A_1^d=B_1^d\). By the induction hypothesis, \(A_0=B_0\) and
	\(A_1=B_1\), so \(A=B\). The weak-negation cases are similar. If \(A=\aneg A_0\), then \(A^d=\uneg A_0^d\). So \(B^d\) has outermost connective \(\uneg\),
	which is possible only if \(B=\aneg B_0\). Then \(A_0^d=B_0^d\), and the
	induction hypothesis gives \(A_0=B_0\). The case \(A=\uneg A_0\) is
	dual.

	The last is the cases in which \(A^d\) has outermost connective
	\(\land\). This can happen in two ways: \(A\) may be a disjunction, or
	\(A\) may be an implication. These two possibilities cannot be confused.
	If \(A=A_0\lor A_1\), then the left conjunct of \(A^d\) is \(A_0^d\),
	which has no main \(\to\), by the first part of the proof. If
	\(A=A_0\to A_1\), then the left conjunct of \(A^d\) is
	\(A_0^d\to\aneg A_0^d\), which has outermost connective \(\to\).

	Therefore, if \(A\) and \(B\) are both disjunctions, say
	\(A=A_0\lor A_1\) and \(B=B_0\lor B_1\), then equality of their
	translations gives \(A_0^d=B_0^d\) and \(A_1^d=B_1^d\). So, we get
	\(A=B\) by the induction hypothesis. If \(A\) and \(B\) are both
	implications, say \(A=A_0\to A_1\) and \(B=B_0\to B_1\), then equality
	of their translations gives \(A_0^d\to\aneg A_0^d =
	B_0^d\to\aneg B_0^d\) and \(A_1^d=B_1^d\). The first equality gives
	\(A_0^d=B_0^d\), and hence \(A_0=B_0\) by the induction hypothesis; the
	second gives \(A_1=B_1\). Thus \(A=B\). All cases have been covered, so the map \(A\mapsto A^d\) is injective.
\end{proof}

\begin{lemma}[Dual valuations]\label{lem:dual-valuations}
	For every \(\mathcal M_{\CLauN}\)-valuation \(v\), there are
	\(\mathcal M_{\CLauN}\)-valuations \(v_1\) and \(v_2\) such that
	\(v_1(A^d)=1-v(A)\) and \(v_2(A)=1-v(A^d)\), for every
	\(A\in\Forms_{\mathcal L_{au}}\).
\end{lemma}

\begin{proof}
	We construct the two valuations separately. First define \(v_2(A)=1-v(A^d)\). We check, by induction on \(A\), that \(v_2\) respects the matrix clauses. The Boolean cases are just the usual De Morgan cases, since the dual translation exchanges \(\land\) and \(\lor\). For implication, observe first that
	\(v(B^d\to\aneg B^d)=1-v(B^d)\): if \(v(B^d)=1\), then
	\(v(\aneg B^d)=0\), so the implication is false. If \(v(B^d)=0\), the
	implication is true. Hence the dual of \(B\to C\), namely
	\((B^d\to\aneg B^d)\land C^d\), has value
	\((1-v(B^d))\land v(C^d)\). Taking the complement gives exactly the
	classical value of \(B\to C\) under \(v_2\).

	It remains to check the weak negations. If \(A=\aneg B\), then
	\(v_2(\aneg B)=1-v(\uneg B^d)\). When \(v_2(B)=1\), we have
	\(v(B^d)=0\), so \(v(\uneg B^d)=1\), and therefore
	\(v_2(\aneg B)=0\), as required. When \(v_2(B)=0\), both values are
	allowed for \(\aneg B\). The case \(A=\uneg B\) is similar: if
	\(v_2(B)=0\), then \(v(B^d)=1\), so \(v(\aneg B^d)=0\), and hence
	\(v_2(\uneg B)=1\); if \(v_2(B)=1\), both values are allowed. Thus
	\(v_2\) is an \(\mathcal M_{\CLauN}\)-valuation.

	We now construct \(v_1\). Let \(S\) contain all formulas of the forms
	\(A^d\) and \(A^d\to\aneg A^d\). This set is subformula-closed: the
	subformulas of a dual formula are again dual formulas or formulas of the
	second form, and \(\aneg A^d\) itself is \((\uneg A)^d\). Define a
	partial assignment \(h\) on \(S\) by putting \(h(A^d)=1-v(A)\) and
	\(h(A^d\to\aneg A^d)=v(A)\). This is well-defined by the preceding
	syntactic lemma: different formulas have different duals, no dual formula
	has principal connective \(\to\), and so no formula \(A^d\) can coincide
	with a formula \(B^d\to\aneg B^d\).

	We check that \(h\) respects the matrix clauses on \(S\). The Boolean
	clauses follow directly from De Morgan duality. For instance, the value
	assigned to \((B\lor C)^d\) is \(1-v(B\lor C)\), which is the conjunction
	of \(1-v(B)\) and \(1-v(C)\), exactly the values assigned to \(B^d\) and
	\(C^d\). The case of \(B\land C\) is dual.

	For implication, \((B\to C)^d\) is
	\((B^d\to\aneg B^d)\land C^d\). By definition, \(h\) gives
	\(B^d\to\aneg B^d\) the value \(v(B)\), and gives \(C^d\) the value
	\(1-v(C)\). Their conjunction is therefore \(v(B)\land(1-v(C))\), which
	is exactly \(1-v(B\to C)\), as required.

	For the weak negations, consider first \(\uneg B^d\), which is
	\((\aneg B)^d\). The assignment gives it the value \(1-v(\aneg B)\). If
	\(h(B^d)=0\), then \(v(B)=1\), so \(v(\aneg B)=0\), and hence
	\(h(\uneg B^d)=1\), as the table for \(\uneg\) requires. If
	\(h(B^d)=1\), the table for \(\uneg\) allows both values. The check for
	\(\aneg B^d\) is dual. Indeed, if \(h(B^d)=1\), then \(v(B)=0\), so
	\(v(\uneg B)=1\), and therefore \(h(\aneg B^d)=0\), as required by the
	table for \(\aneg\); if \(h(B^d)=0\), both values are allowed.

	Finally, we check the formulas \(A^d\to\aneg A^d\). If \(v(A)=1\), then
	\(h(A^d)=0\), so the implication is true. If \(v(A)=0\), then
	\(h(A^d)=1\), and the table for \(\uneg\) forces \(v(\uneg A)=1\). Hence
	\(h(\aneg A^d)=h((\uneg A)^d)=0\), so the implication is false. Thus
	\(h(A^d\to\aneg A^d)=v(A)\), exactly as stipulated.

	So \(h\) is a  partial valuation on a subformula-closed set. Since
	all matrix operations return non-empty sets of values, we can extend \(h\)
	recursively to all formulas by making arbitrary appropriate choices outside
	\(S\). The resulting total valuation is \(v_1\), and by construction
	\(v_1(A^d)=1-v(A)\) for every \(A\).
\end{proof}

\begin{theorem}[Duality]\label{thm:duality}
	For all \(\Gamma,\Pi\subseteq\Forms_{\mathcal L_{au}}\),
	\[
		\Gamma\Vdash_{\CLauN}\Pi
		\quad\text{iff}\quad
		\Pi^d\Vdash_{\CLauN}\Gamma^d.
	\]
\end{theorem}

\begin{proof}
	We argue contrapositively. Suppose first that
	\(\Pi^d\not\Vdash_{\CLauN}\Gamma^d\). Then some valuation \(v\) makes
	every \(P^d\), with \(P\in\Pi\), true and every \(G^d\), with
	\(G\in\Gamma\), false. By Lemma~\ref{lem:dual-valuations}, there is a
	valuation \(v_2\) such that \(v_2(A)=1-v(A^d)\) for every \(A\). Hence
	\(v_2\) makes every member of \(\Gamma\) true and every member of
	\(\Pi\) false. Thus \(\Gamma\not\Vdash_{\CLauN}\Pi\).
	Conversely, suppose that \(\Gamma\not\Vdash_{\CLauN}\Pi\). Then some
	valuation \(v\) makes every member of \(\Gamma\) true and every member
	of \(\Pi\) false. By Lemma~\ref{lem:dual-valuations}, there is a
	valuation \(v_1\) such that \(v_1(A^d)=1-v(A)\) for every \(A\). Hence
	\(v_1\) makes every member of \(\Pi^d\) true and every member of
	\(\Gamma^d\) false. Thus \(\Pi^d\not\Vdash_{\CLauN}\Gamma^d\).
\end{proof}



\section{Conclusions}\label{sec:Concl}

We introduced three two-negations paraclassical logics, $\CLoaN$, $\CLouN$,
and $\CLauN$, each with a sound and complete two-valued non-deterministic
semantics, and studied $\CLauN$ in detail. There, classical negation is
table-definable by $p\to\aneg p$, the negation $\oneg$ is definable, and a
boundary parameter $\sim_o^{au}$ controls the two weak negations, arranging
all four into the negation diamond and yielding the decompositions $\aneg
	A\equiv\cneg A\land\sim_o^{au}A$ and $\uneg A\equiv\cneg A\lor\sim_o^{au}A$.
Using definable consistency and determinacy operators, which place $\CLauN$
near the logics of formal inconsistency \cite{Carnielli2007}, we recaptured
classical logic both locally and via a translation that encodes the boundary
parameter as fresh atoms.
Among the possible directions of future work, we mention three: a cut-free proof system for $\CLauN$ (and
thence decidability and complexity), the extent to which the expressivity
results carry over to $\CLoaN$ and $\CLouN$, and whether Craig interpolation
transfers to $\CLauN$ through its translation into $\CL$.

\paragraph{Use of AI and FUNDING}
Generative AI tools were used to assist with linguistic editing, LaTeX troubleshooting, and exploratory brainstorming about possible proof strategies and conceptual connections between the systems. All final formulations, mathematical arguments, examples, references, and claims were independently checked by the authors.
Pawel Pawlowski has been supported by the Research Foundation – Flanders (FWO) under grant no.~ 1255724N (Senior Postdoc Fellowship). Pietro Vigiani acknowledges that the first submitted version of this article has been produced with the financial support of the Research Foundation – Flanders (FWO) under grant no.~1215326N  (Junior Postdoc Fellowship); and that the revised version of this article has been produced with the financial support of the European Union under the project: Biography of Fake News with a Touch of AI: Dangerous Phenomenon through the Prism of Modern Human Sciences project no.: CZ.02.01.01/00/23\_025/0008724 via the Operational Programme Jan Amos Komenský.
\bibliographystyle{eptcs}
\bibliography{haptism}

\end{document}